%% file: main.tex
\documentclass[12pt]{iopart}

\usepackage{xcolor}
\usepackage{graphicx}
\usepackage{float} 
\usepackage{caption}

\begin{document}

\title{Beyond $\Lambda$CDM: How the Hubble tension challenges early universe physics}

\author{Gawain Simpson, Krzysztof Bolejko, Stephen Walters}

\address{School of Natural Sciences, College of Sciences and Engineering, University of Tasmania, Private Bag 37, Hobart TAS 7001, Australia}
\ead{gawain.simpson@utas.edu.au}

\begin{abstract}
Differences in the values of the Hubble constant obtained from the local universe and the early universe have resulted in a significant tension. This tension signifies that our understanding of cosmology (physical processes and/or cosmological data) is incomplete. Some of the suggested solutions include physics of the early Universe. 
In this paper we aim to investigate common features of various early-universe solutions to the Hubble constant tension. 
The physics of the early universe affects the size of the sound horizon which is probed with the Cosmic Microwave Background (CMB) data. Within the standard model, the size of the horizon (within limits of current measurements) is affected by processes that could occur between (approximately) 1 day after the Big Bang and the last scattering instant. 
We focus on simple extensions incorporating Early Dark Energy (EDE) and show how such a model affects the inferred values of the Hubble constant. We compare this model to $\Lambda$CDM models using MCMC analysis, likelihoods over the parameter space and Bayesian evidence.
The MCMC analysis shows that EDE leads to a decrease in the size of the sound horizon that is consistent with $H_0=73.56$~km~s$^{-1}$~Mpc$^{-1}$ but we also show that MCMC analysis favours increasing redshift and proportion of EDE. The Bayesian evidence favours our EDE model for very narrow, finely-tuned parameter space. The $\Lambda$CDM model used for comparison has good evidence across a wide parameter space. We interpret this as an indication that more sophisticated models are required.
We conclude that if the Hubble tension were to be related to the physics of the early universe, EDE could be used as a window to explore conditions of the early universe and extend our understanding of that era.
\end{abstract}

\section{Introduction}\label{intro}
Georges Lema\^itre and Edwin Hubble discovered the expansion of the universe in 1927 and 1929 respectively, and each derived an expansion rate of the universe which has been called the Hubble constant. While the value has since been revised down significantly, there is now some disagreement, or tension, in the current measured value. There are two main methods of measuring the Hubble constant. The first one relates to the physics of the late universe, the second is derived from parameters of the Cosmic Microwave Background (CMB). The current values are of the order of 73 and 67 km s$^{-1}$ Mpc$^{-1}$ respectively, and the uncertainties in the values are sufficiently small that there is a significant tension between them.

In 2023, Kamionkowski and Riess \cite{2023ARNPS..73..153K} conducted a thorough review of research published into the Hubble Tension. They examined the values of the Hubble constant obtained from observations of the local universe by multiple methods and compared those to values calculated from measurements of the early universe, followed by a review of early- and late-universe models that may resolve the Tension. 
They concluded that Early Dark Energy (EDE) may be the most likely candidate to resolve the Hubble Tension. 

In this paper, we posit a hypothesis that the Hubble tension could be due to some process in the early universe that is not yet incorporated into the standard model. We use this hypothesis as an opportunity to investigate possible extensions and properties of the early universe. We constructed a model incorporating EDE and used a Markov Chain Monte Carlo (MCMC) analysis to see if it can resolve the Hubble tension. We reviewed our results by examining the likelihood across the parameter space of the MCMC analysis and also the Bayesian evidence for our model. It is anticipated that within the next decade, with data coming from next generation surveys, this hypothesis will be thoroughly tested and either rejected or will indeed provide a window to the physics of the early universe.

The remainder of this paper is structured as follows: Section 1.1 finishes the introduction with a brief review of EDE and early universe models to position our paper within this field. Section 2 discusses the sound horizon, how it affects the Hubble constant and describes the research in this paper to resolve the Hubble tension. Section 3 gives the results of our research and discussion of these and section 4 contains our conclusions.

\subsection{Review of EDE and other early universe models}\label{review}
Here we look at recent research into EDE models, followed by other early universe models, each aimed to reduce the Hubble tension.

The EDE is a phenomenological umbrella term that encompasses a wide range of physical models.
A number of teams have investigated EDE models and its possible properties that would allow us to identify what type of phenomena EDE could be related to.
Beanoum et al. \cite{2023arXiv230705917B} presented an EDE model featuring a non-linear electrodynamics framework that mimics radiation at the early stages of the Universe and exhibits an accelerated expansion in the late evolution of the Universe.
Cruz et al. \cite{2023PhRvD.108b3518C} developed a New Early Dark Energy (NEDE) model, with a vacuum energy component decaying around recombination, providing a solution to the Hubble tension, comparable to the Supernovae and H0 for the Equation of State of dark energy program (SH0ES) \cite{2022ApJ...934L...7R} measurement of $H_0$.
Herold and Ferreira \cite{2023PhRvD.108d3513H} presented constraints on the EDE fraction and the Hubble parameter, endorsing EDE as a viable solution to the Hubble tension.
Lin, McDonough, Hill and Hu \cite{2023PhRvD.107j3523L} investigated an Early Dark Sector (EDS) where the mass of dark matter depends on the EDE scalar field. Their model, termed trigger EDS, performed comparably to EDE in resolving the Hubble tension.
Niedermann and Sloth \cite{2023arXiv230703481N} argued that NEDE offers a natural framework to extend $\Lambda$CDM in order to address the $H_0$ tension. They discussed triggers at a microscopic level, either an ultralight scalar field or a dark sector temperature.

Poulin, Smith and Karwal \cite{2023arXiv230209032P} reviewed various EDE models, emphasizing their potential to resolve the Hubble tension, with next-generation surveys as crucial testing grounds. In a further paper \cite{2023PDU....4201348P} they conducted a thorough review of EDE solutions to the Hubble Tension. They reviewed eight different significant EDE models, and some minor models, and concluded that EDE endures as a solution to the Hubble tension but does not definitively solve the tension, which will require data from future observatories to provide more precise measurements. They also discussed the research showing that EDE models increase the $S_8$ tension. While we acknowledge this, it is outside the scope of this paper.

Other groups examined scenarios not limited to EDE.
Investigating a coupling between neutrinos and an EDE scalar field, de Souza and Rosenfeld \cite{2023arXiv230204644D} explored a potential easing of the Hubble tension. However, their parameters did not yield values relieving the tension. 
Gonzalez et al. \cite{2023arXiv230209091C} examined the claims of de Souza and Rosenfeld \cite{2023arXiv230204644D}, disagreeing with the findings and asserting that neutrino-assisted EDE remains an interesting potential resolution of the Hubble tension, advocating further study.
Escamilla et al. \cite{2023arXiv230714802E} examined the Dark Energy (DE) equation of state (EoS), finding no compelling evidence for inconsistency with $w = -1$.

Research has also been conducted into axion-like EDE Models. Cicoli et al. \cite{2023JHEP...06..052C} modelled EDE in type IIB string theory with the EDE field identified as either a C4 or C2 axion. In a different approach Eskilt et al. \cite{2023arXiv230315369E} examined polarization of the cosmic microwave background (CMB), connecting it to axion-like EDE and its potential violation of parity symmetry. Goldstein et al. \cite{2023arXiv230300746G} presented constraints on axion-like EDE using Lyman-$\alpha$ absorption lines, suggesting that severe constraints on EDE models could resolve the Hubble tension, and Jiang and Piao \cite{2022PhRvD.105j3514J} investigated axion-like EDE, combining CMB observations to favour non-zero EDE fractions and large Hubble constants.

Another approach to resolve the Hubble tension was to modify $\Lambda$CDM itself.
Lemos et al. \cite{2023EPJC...83..495L} tested the consistency of $\Lambda$CDM assumptions at z~$\sim$~1000, emphasizing the robustness of the model but also the need for more precise measurements.
Lee et al. \cite{2023PhRvL.130p1003L} explored data-driven solutions to the Hubble tension via perturbative $\Lambda$CDM cosmology modifications, using the Fisher bias formalism. Focusing on a dynamic electron mass and fine structure constant, they found that modifying recombination, based on Planck CMB data, could potentially resolve the tension. However, inclusion of BAO and uncalibrated supernovae data tempered this optimism. 
Wang and Piao \cite{2022PhLB..83237244W} explored varying equations of state in $\Lambda$CDM through MCMC analysis, favouring only marginal evolution in $w$, other than the typical $w=-1$. In a separate study \cite{2022arXiv220909685W}, they investigated coupling a fraction of dark matter with EDE, finding it alleviated the S8 tension but led to a diminished fit for $H_0$.

The latest data release (DR6) from the Atacama Cosmology Telescope shows essentially no shift in $H_0$ within the EDE model, compared to $\Lambda$CDM, with the hint of non-zero EDE, which was seen in data release 4, absent in DR6 \cite{2025arXiv250314454C}.

Takahashi and Toda \cite{2023arXiv230600454T} investigated the impact of big bang nucleosynthesis, revealing that its analysis could alter the tension by 0.8$\sigma$ in EDE models. 
Meanwhile, Tian and Zhu \cite{2023PhRvD.107j3507T} proposed fluid property modifications, inspired by the radiation-matter transition, avoiding the need for fluid-like dark energy.

Ye, Jiang, and Piao \cite{2023arXiv230518873Y} used CMB-only data to constrain the shape of lensing in EDE models, irrespective of late Universe evolution. They explored how CMB data, including temperature, polarization spectra, and lensing reconstruction, restricts the lensing potential in $\Lambda$CDM and EDE. Current CMB lensing data enforces a stringent constraint in the 80~$<l<$~400 range of the CMB power spectrum, requiring $\Lambda$CDM-like behavior. This imposes a robust constraint on late Universe behaviors during the relevant CMB lensing period.

EDE could also be a signature not of the actual physical field/process, but a manifestation of limitation of our understanding of gravity. In the area of modifications to general relativity and related cosmological models, 
Odintsov et al. \cite{2023PhLB..84337988O, 2021NuPhB.96615377O} explored a power-law F(R) gravity scenario with an EDE term. The EDE term, acting between matter-radiation equality and recombination, did not significantly impact $H_0$ values but required a substantial coefficient for viability. 
Ben-Dayan and Kumar \cite{2023arXiv230200067B} investigated Banks-Zaks/Unparticles as a dark energy model, obtaining $H_0$~$\approx$~70–73 km s$^{-1}$ Mpc$^{-1}$.
Hu and Wang \cite{2023Univ....9...94H} reviewed the Hubble tension, advocating for late universe modification. 
Khodadi and Schreck's models \cite{2023PDU....3901170K}, based on modified General Relativity, failed to explain the Hubble tension, highlighting limitations of new physics. Kuzmichev and Kuzmichev \cite{2022arXiv221116394K} proposed a quantum Bohm potential-inspired term as a candidate for addressing the Hubble tension, altering the early universe's expansion history while leaving the late evolution unchanged.

Vagnozzi proposed that early universe solutions alone can not solve the Hubble tension and a combined approach is needed, requiring modifications to early time, late time and local physics \cite{2023Univ....9..393V}.
Akarsu et al. \cite{2024Univ...10..305A} explored the emerging discrepancies between predictions from $\Lambda$CDM and observations and suggested that tensions hint at missing physics in the late Universe and future observations will help clarify whether the tensions are due to problems in $\Lambda$CDM or from observational variability.

In summary, no single theory or modification explored above could resolve the Hubble tension and any solution will likely consist of both early and late universe modifications. The research reviewed concluded that EDE may contribute but will likely only be a part of the solution.

\section{The Sound Horizon and the Hubble Constant}\label{The Sound Horizon}

The pre-CMB universe consisted primarily of a plasma of baryons, photons and other particles such as neutrinos. Interactions between photons and baryons in this fluid created perturbations in the plasma which propagated at the speed of sound. At recombination, matter and radiation decoupled and the fluctuations are now visible as primary: CMB anisotropies. The largest of these correspond to the sound horizon which was the distance sound waves in the photon-baryon fluid could travel just before recombination.

\hfill \break
The sound horizon, $r_s$ at the CMB is given by
\begin{equation}\label{eqn r_s}
(1+z){r_s} = \int_{0}^{t_{CMB}} \frac{ c_s (t) } {a(t)} {\rm d}t,
\end{equation}
where $z$ is the redshift, $t_{CMB}$ is the age of the universe at the CMB and $a$ is the scale factor.
The speed of sound $c_s$ is given by
\begin{equation}\label{eqn c_s}
c_s^2 =\frac{1}{3} \left( 1+\frac{3\, \omega_b}{4\,\omega_{\gamma} (1+z)} \right)^{-1},
\end{equation}
where $\omega_b$ is the baryon density and $\omega_\gamma$ is the photon density.
The Hubble constant $H_0$ is related to the scale factor $a$ by
\begin{equation}\label{eqn H0_a}
H_0 =\frac{\dot a_0}{a_0}.
\end{equation}

\begin{figure}[H]
    \centering
    \captionsetup{justification=justified, margin=1.5cm, font=small}
    \includegraphics[width=14cm]{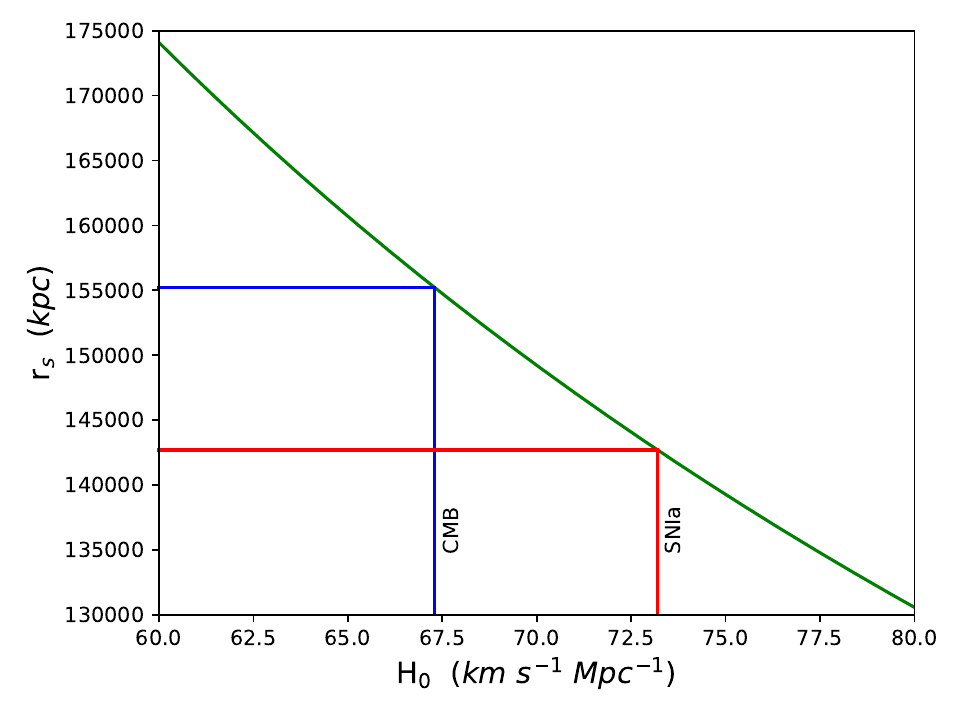}
    \caption{Size of the sound horizon $r_s$ for values of the Hubble constant $H_0$, using equation \ref{eqn r_s}. Values of $r_s$ for values of $H_0$ from the CMB ($H_0=67.3$~km~s$^{-1}$~Mpc$^{-1}$) \cite{2020A&A...641A...6P} and SNIa observations ($H_0=73.3$~km~s$^{-1}$~Mpc$^{-1}$) \cite{2022ApJ...934L...7R} are indicated.}
    \label{fig:r_s vs H0}
\end{figure}

Equations (\ref{eqn r_s}) and (\ref{eqn H0_a}) show the size of the sound horizon is dependent on the Hubble constant, shown in Figure \ref{fig:r_s vs H0}. Indicated on the Figure is the size of the sound horizon for values of the Hubble Constant from the values consistent with early and late universe data (keeping $\Omega_M, \Omega_R$ and $\Omega_\Lambda$ fixed).

Figure \ref{fig:r_s vs H0} is a simple illustration of various physical models briefly reviewed in Section~\ref{intro}.
The aim of introducing new physics process into the standard model of the early universe aims to decrease the size of the sound horizon at the CMB.
Below, we review standard equations and investigate how various terms affect the size of the sound horizon. 

\subsection{Decreasing the sound horizon}

\begin{figure}[H]
    \centering
    \captionsetup{justification=justified, margin=2cm, font=small}    
    \includegraphics[width=12cm]{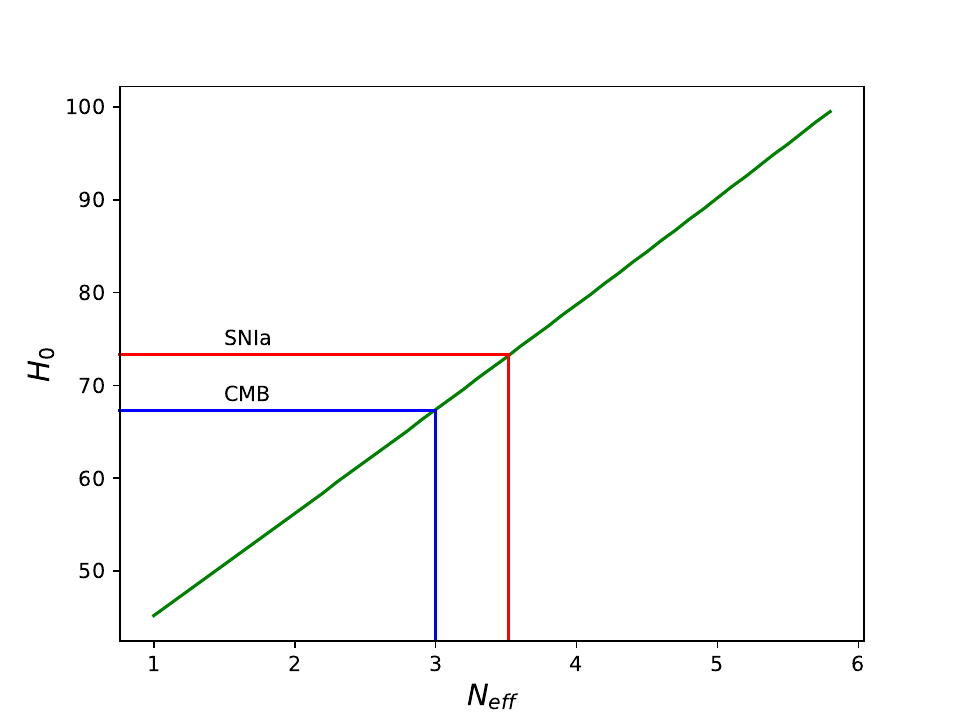}
    \caption{The inferred value of the Hubble constant from the CMB and its dependence on the effective number of neutrino species, $N_{eff}$, showing a linear relationship between $H_0$ and $N_{eff}$. Values of $N_{eff}$ for values of $H_0$ from the CMB ($H_0=67.3$~km~s$^{-1}$~Mpc$^{-1}$) \cite{2020A&A...641A...6P} and SNIa observations ($H_0=73.3$~km~s$^{-1}$~Mpc$^{-1}$) \cite{2022ApJ...934L...7R} are indicated.}
    \label{fig:H0 vs Neff}
\end{figure}

The early universe was radiation dominated. In the standard model, the radiation is sourced by two components: photons and neutrinos. 
The parameter $\Omega_R = \omega_r/h^2$ (where $h = H_0/100$), and the equation for the radiation density, $\omega_r$ is given by:
\begin{equation}\label{Neff}
\omega_r = \left[1 + \frac{7}{8}N_{eff} \left(\frac{4}{11}\right)^{\frac{4}{3}}  \right]\omega_{\gamma},
\end{equation}
where $\omega_{\gamma}$ is the photon density and $N_{eff}$ is the effective number of neutrino species in the early universe. It can be seen that increasing the number of neutrino species increases the Hubble constant, shown in Figure \ref{fig:H0 vs Neff} for $\Omega_M=0.31, \Omega_R=0.0001$ and $\Omega_\Lambda=0.69$. The value used for the remainder of this paper is $N_{eff}=3.046$ \cite{2020A&A...641A...6P}.

\hfill \break
The sound horizon is derived from the angular diameter distance to the CMB, $\theta_s$, which has been measured by the Planck space observatory to a high degree of accuracy, with a value of $\theta_s=(1.04108 \pm 0.00031)\times 10^{-2}$ \cite{2020A&A...641A...6P}, with $\theta_s = \frac{r_s}{D_A}$, where $r_s$ is the  size of the CMB sound horizon and $D_A$ is the angular diameter distance to the CMB.

The equation for $D_A$ is also derived from parameters of the CMB and given by:
\begin{equation}\label{eqn D_A}
(1+z){D_A} = \frac{c}{H_0}\int_{0}^{z_{_{CMB}}} \frac{dz}{(\Omega_M(1+z)^3 + \Omega_R(1+z)^4 + \Omega_{\Lambda})^{\frac{1}{2}}}.
\end{equation}

\begin{figure}[H]
    \centering
    \captionsetup{justification=justified, margin=2cm, font=small}
    \includegraphics[width=14cm]{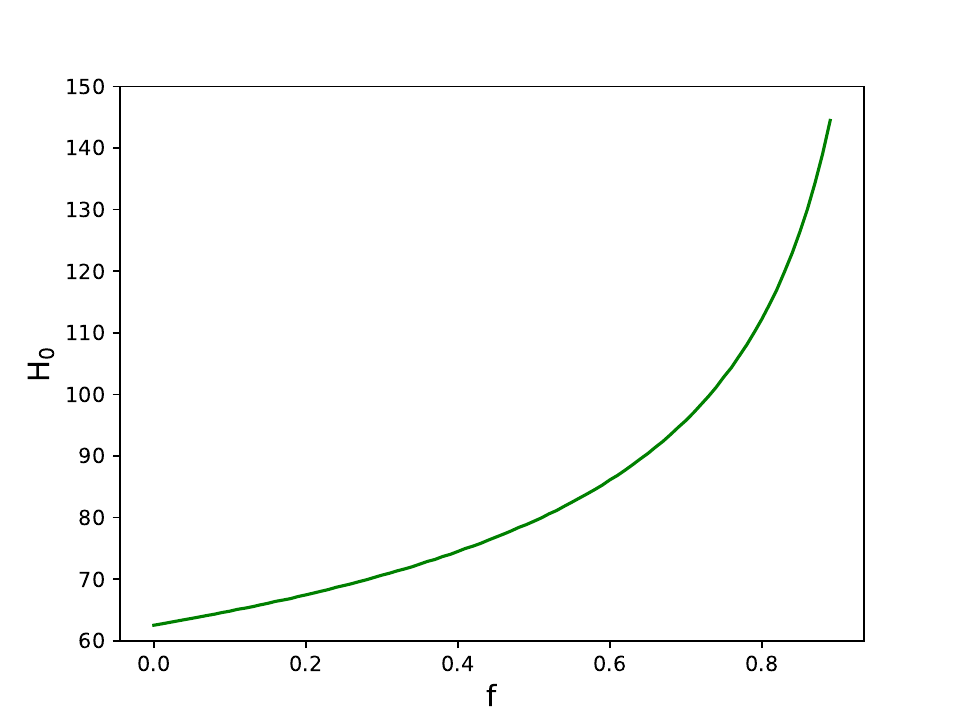}
    \caption{The predicted Hubble constant $H_0$ increases with an increasing proportion $f$ of Early Dark Energy as a fraction of the total energy of the
    early universe.}
    \label{fig:H0 vs f}
\end{figure}

\hfill \break
In this paper, we propose to introduce Early Dark Energy (EDE) into the pre-CMB universe and study the effect it has on altering the size of the sound horizon and therefore changing the value of the Hubble constant in the present universe. We modeled this by adding an additional term, EDE, to the equation for the sound horizon: 
\begin{equation}\label{eqn r_s_EDE}
(1+z){r_s} = \frac{c}{H_0}\int_{z_{_{CMB}}}^{\infty} \frac{dz}{(\Omega_M(1+z)^3 + \Omega_R(1+z)^4 + \Omega_{\Lambda} + {\Omega_{EDE}})^{\frac{1}{2}}}\frac{1}{\sqrt{3[1+\frac{3\, \omega_b}{4\,\omega_{\gamma} (1+z)}]}},
\end{equation}
where {$\Omega_{EDE}$} is the additional EDE component that affects the size of the sound horizon and is given by:

\begin{equation}\label{eqn EDE}
\Omega_{EDE} = \frac{2}{{\left({\frac{1+z_c}{1+z}}\right)^{3(1+w_n)}+1}}\,\frac{f}{1-f}(\Omega_M(1+z_c)^3 + \Omega_R(1+z_c)^4 + \Omega_{\Lambda}),
\end{equation}
where $z_c$ is the critical redshift at which EDE is at its maximum and $f$ is the fraction of total energy which is EDE at $z_c$. Increasing EDE as a fraction of total energy of the early universe, i.e. increasing f, increases $H_0$ as shown in Figure \ref{fig:H0 vs f}, for $z_c=20,000$ and $\Omega_M=0.3147$ \cite{2020A&A...641A...6P} .

\begin{figure}[H]
    \centering
    \captionsetup{justification=justified, margin=1.5cm, font=small}
    \includegraphics[width=14cm]{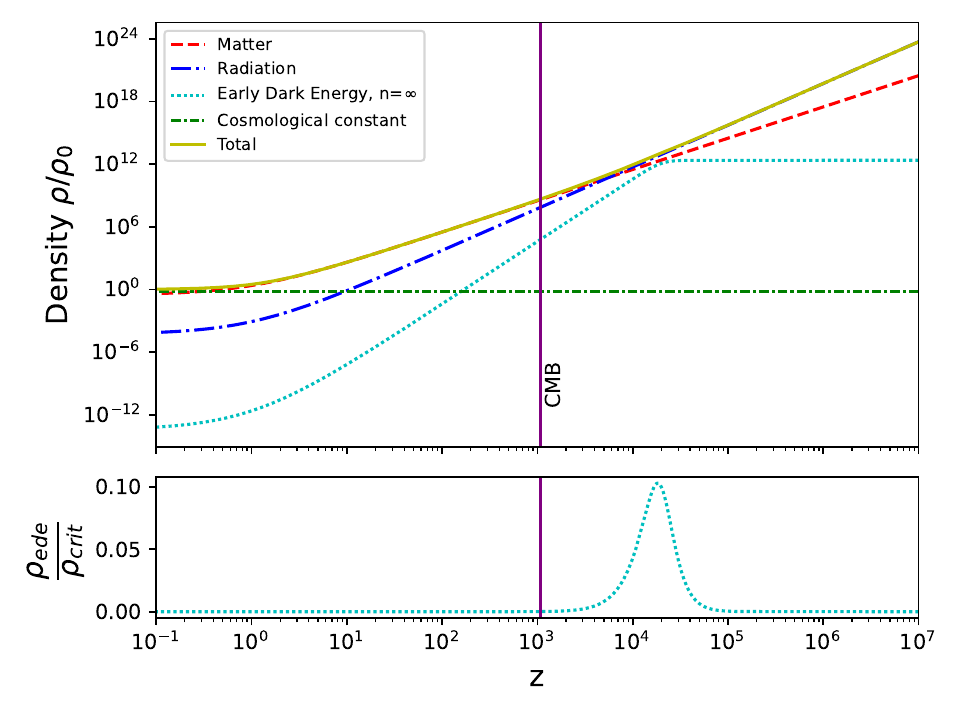}
    \caption{Evolution of Densities of Cosmological Fluids, including Early Dark Energy, from z = $10^7$ to $10^{-1}$ (R to L). The upper plot shows the densities of matter, radiation, EDE, the cosmological constant and a total of these as a proportion of current universe density. The lower plot shows the EDE density as a proportion of critical density. The vertical line in the centre marks the redshift of the CMB.}
    \label{fig:evol_ede}
\end{figure}

The expanding universe causes different cosmological fluids to evolve differently.
Figure \ref{fig:evol_ede} shows the evolution of all cosmological fluids, Matter, Radiation, the Cosmological constant (Dark Energy) and EDE, as a function of redshift. The lower plot shows EDE as a proportion of total energy density and how it peaks at a critical redshift, for $f=0.1$ and $z_c=20,000$. Matter reduces $\propto a^{-3}$, radiation $\propto a^{-4}$, dark energy (cosmological constant) $\propto a^0$ and EDE peaks in the early universe, then fades away, becoming essentially insignificant by the CMB.

The Nine-year Wilkinson Microwave Anisotropy Probe (WMAP) Observations \cite{2013ApJS..208...19H} measured the power spectrum of the cosmic microwave background radiation temperature anisotropy to significant accuracy, yielding measurements of the physical baryon density, the physical cold dark matter density and the dark energy density, and derived a value for $H_0 = 69.33 \pm 0.88$ km s$^{-1}$ Mpc$^{-1}$. This is in tension with the value of $H_0$ derived from SNIa observations, which is $H_0=73.30\pm1.04$ km s$^{-1}$ Mpc$^{-1}$ \cite{2022ApJ...934L...7R}.

\begin{figure}[H]
    \centering
    \captionsetup{justification=justified, margin=2cm, font=small}
    \includegraphics[width=14cm]{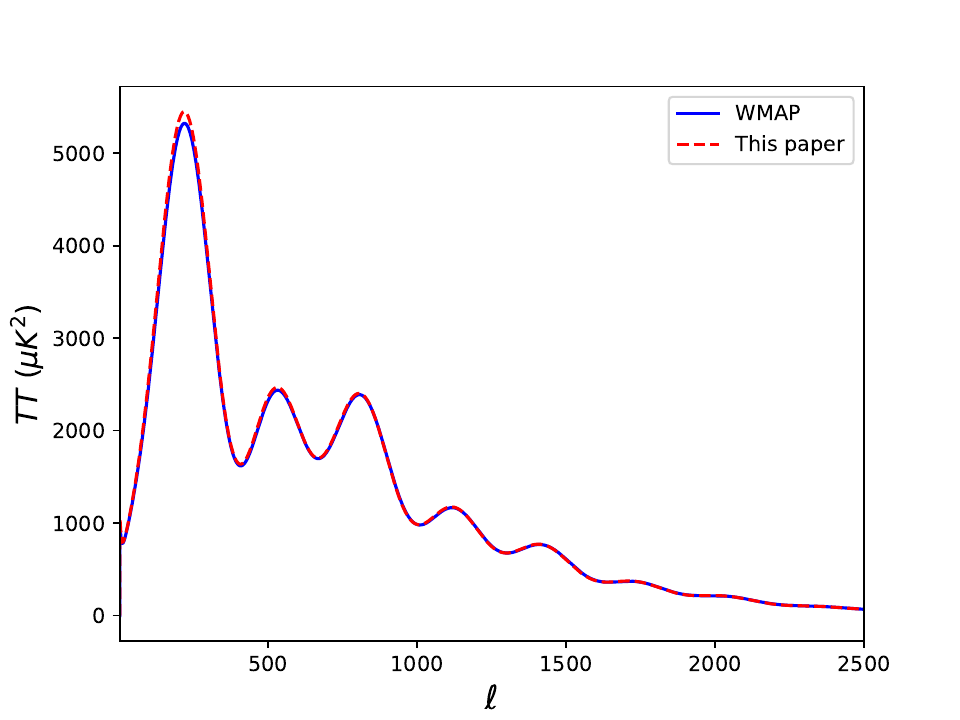}
    \caption{Power spectrum of CMB from WMAP observations and with modified parameters to obtain a similar power spectrum.}
    \label{fig:CAMB}
\end{figure}

In order to see if different parameters could be found to alleviate the Hubble tension, we chose a nominal value for $\Omega_M = 0.27$, maintaining the same value of $\Omega_b h^2$, and found that a value of $H_0 = 72$ km s$^{-1}$ Mpc$^{-1}$ was the best fit to obtain a matching power spectrum, using the CAMB python package \cite{2011ascl.soft02026L}. The WMAP and modified parameter power spectra are shown in Figure \ref{fig:CAMB}, showing that a similar power spectrum to WMAP can be obtained with different parameters, indicating that different conditions of the early universe can result in a similar power spectrum to that observed. The modified parameter values could then be the basis of a pre-CMB scenario which includes EDE.

The aim of the research was to use Markov Chain Monte Carlo (MCMC) analysis to find whether adding EDE to the pre-CMB universe would alter $H_0$ sufficiently to resolve the Hubble tension.

An initial approach was to test whether a simple fluid, rather than the EDE in equation \ref{eqn EDE}, would suffice to resolve the Hubble tension. To this end, we conducted an MCMC analysis using a simple fluid, $\Omega_X$, in place of the EDE, and also a complicated fluid, ${\Omega_X (1+z)}^{3(1+Y)}$. In the first case $\Omega_X$ had an initial value of 0.5 and was allowed to vary between 0 and 1. For the second case $\Omega_X$ had an initial value of 0.1 and was allowed to vary between 0 and 1, while Y had an initial value of 1 and was allowed to vary freely.

An MCMC analysis was then performed using the EDE in equation \ref{eqn EDE} to find the best fit values of the critical redshift $z_c$ and EDE fraction $f$, using the python emcee package \cite{2013PASP..125..306F} for these values of $\Omega_M$ and $H_0$, using equations \ref{eqn D_A} to \ref{eqn EDE} and fixing the values of $\theta_s$ and $\omega_b$ to the Planck data \cite{2020A&A...641A...6P}. The values of $\theta_s$ and its uncertainty, along with a calculated $D_A$ were used in the probability of acceptance in the MCMC analysis, with constraints $0<f<1$ and $z_c>1100$. We then compared the results from this to an MCMC analysis based on Planck \cite{2020A&A...641A...6P} and Pantheon \cite{2022ApJ...938..110B} data, and also just on Planck data. Lastly we obtained the Bayesian evidence for each model, using the python Dynesty package \cite{2020MNRAS.493.3132S, 2024zndo..12537467K, 2004AIPC..735..395S, 10.1214/06-BA127, 2009MNRAS.398.1601F} and compared the evidence for each model. Comparison of the evidence for each model balances $\chi$-squared minimization against model complexity, with higher evidence value indicating a model is more favoured when considering both fit and complexity.

\section{Results and Discussion}\label{Results}
\begin{figure}[H]
    \centering
    \captionsetup{justification=justified, margin=1.5cm, font=small}
    \includegraphics[width=12cm]{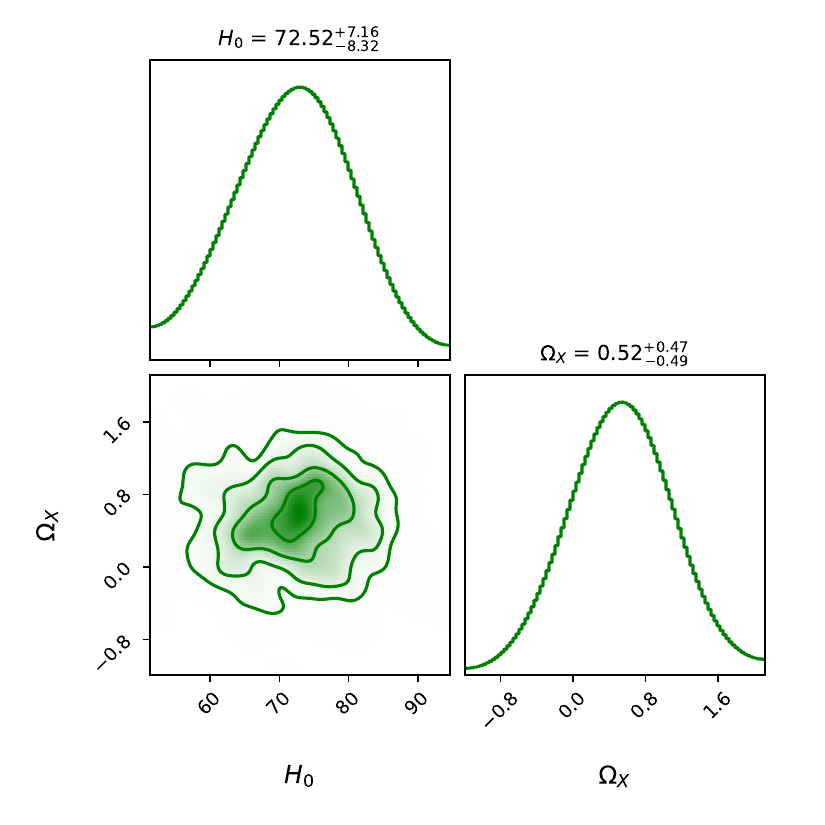}
    \caption{$H_0$ MCMC plot for simple fluid $\Omega_X$, resulting in just Gaussian noise. The median values of $H_0$ and $\Omega_X$ do not show deviation from their initial values. The uncertainties in the value of $\Omega_X$ are of the order of $\Omega_X$ itself and the uncertainties in $H_0$ are too large to indicate any dependence on $\Omega_X$ or be conclusive.}
    \label{fig:omegax}
\end{figure}

The MCMC analysis using simple fluids did not prove useful in resolving the Hubble tension. The addition of the simple fluid $\Omega_X$ results in essentially just Gaussian noise, as shown in Figure \ref{fig:omegax}. The median values of $H_0$ and  $\Omega_X$ have not varied from their initial values and the uncertainties in the value of $\Omega_X$ are of the order of $\Omega_X$ itself and the uncertainties in $H_0$ are too large to indicate any dependence on $\Omega_X$ or be conclusive.

\begin{figure}[H]
    \centering
    \captionsetup{justification=justified, margin=1.5cm, font=small}
    \includegraphics[width=14cm]{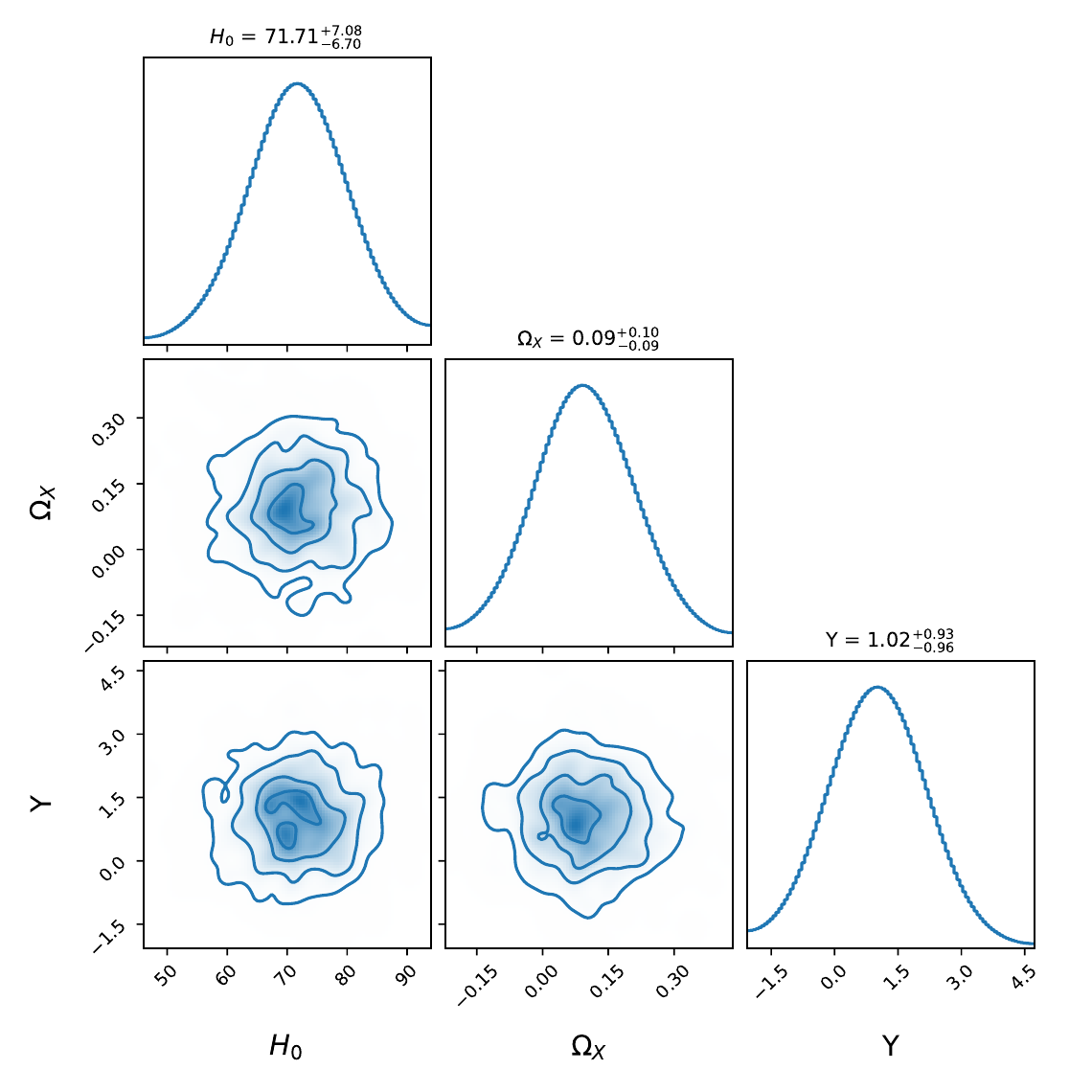}
    \caption{$H_0$ MCMC plot for fluid ${\Omega_X(1+z)}^{3(1+Y)}$, resulting in just Gaussian noise. Similar to Fig. \ref{fig:omegax}, the median values of $H_0$, $\Omega_X$ and Y are of the order of each parameter and the uncertainties in the parameters again inconclusive.}
    \label{fig:omegaxy}
\end{figure}

Similarly, the fluid ${\Omega_X(1+z)}^{3(1+Y)}$ also results in Gaussian noise (Figure \ref{fig:omegaxy}), with the median values of $H_0$, $\Omega_X$ and Y being of the order of each parameter and the uncertainties in $H_0$, $\Omega_X$ and Y again inconclusive. 

\begin{figure}[H]
    \centering
    \captionsetup{justification=justified, margin=1.5cm, font=small}
    \includegraphics[width=15.5cm]{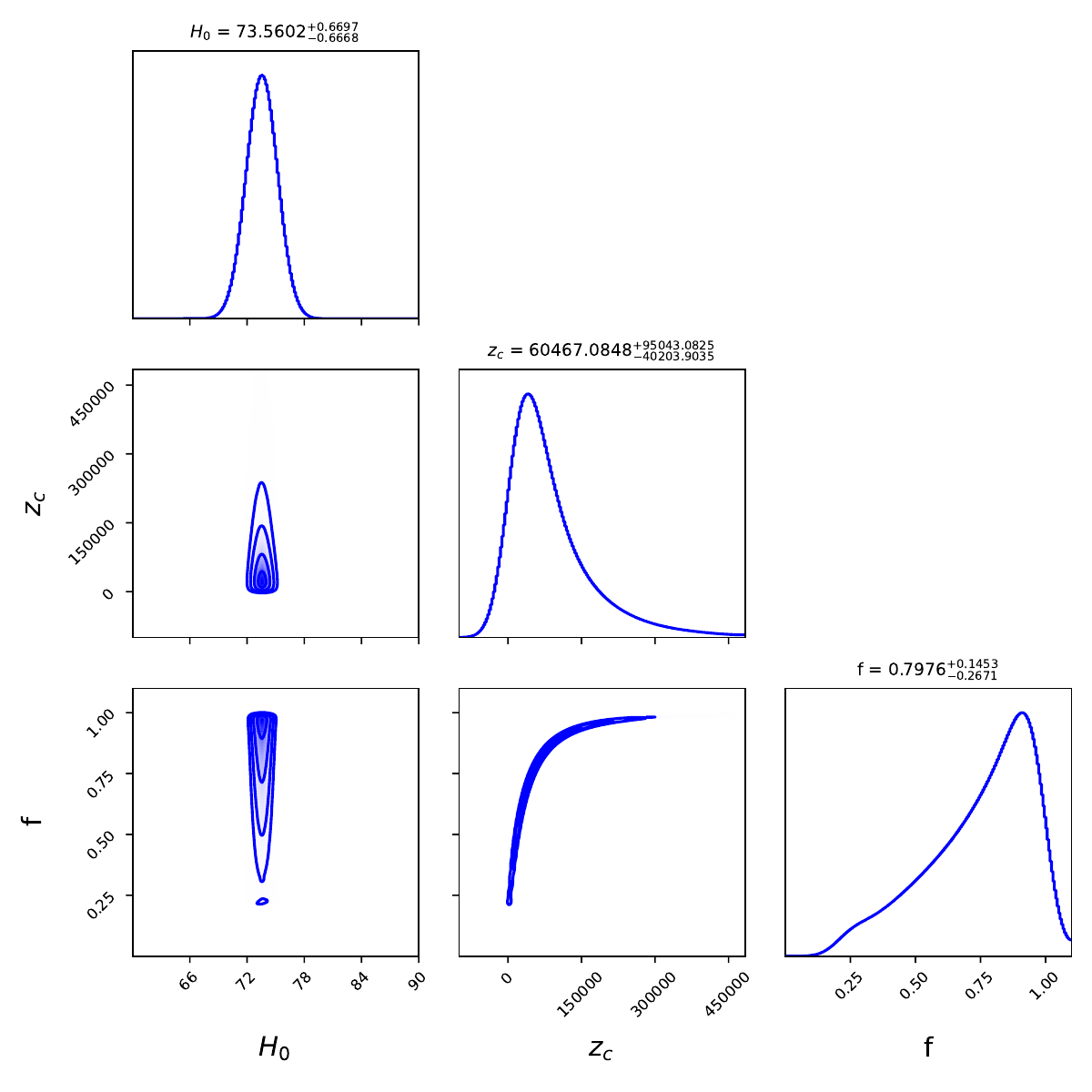}
    \caption{$H_0$ MCMC plot with EDE. A high fraction of EDE, at high redshift, results in $H_0$ = 73.56 km s$^{-1}$ Mpc$^{-1}$, in general agreement with observations from the local universe.}
    \label{fig:mcmc_H0_zc_f}
\end{figure}

Figure \ref{fig:mcmc_H0_zc_f} shows the results of the MCMC analysis for a model incorporating EDE with $\Lambda$CDM. To achieve a value of the Hubble constant that is in agreement with observations from the local universe, such as SH0ES, a fraction as high as $f = 0.798\ ^{+0.15}_{-0.28}$ and a median critical redshift $z_c = 60467 \ ^{+95043}_{-40204}$ was found to be the best fit, resulting in $H_0 = 73.56$ km s$^{-1}$ Mpc$^{-1}$, meaning the peak EDE occurs early and at a large fraction of the total energy of the universe.

The addition of EDE in figure \ref{fig:mcmc_H0_zc_f} indicates that an additional fluid in the pre-CMB universe can alleviate the Hubble tension but the large fraction $f$ at a large and increasing redshift $z_c$ means that a distinct fraction and redshift has not been determined and further work is necessary to determine the precise fraction of the universe that EDE represents and its associated redshift.

\begin{figure}[H]
    \centering
    \captionsetup{justification=justified, margin=1cm, font=small}
    \includegraphics[width=15.5cm]{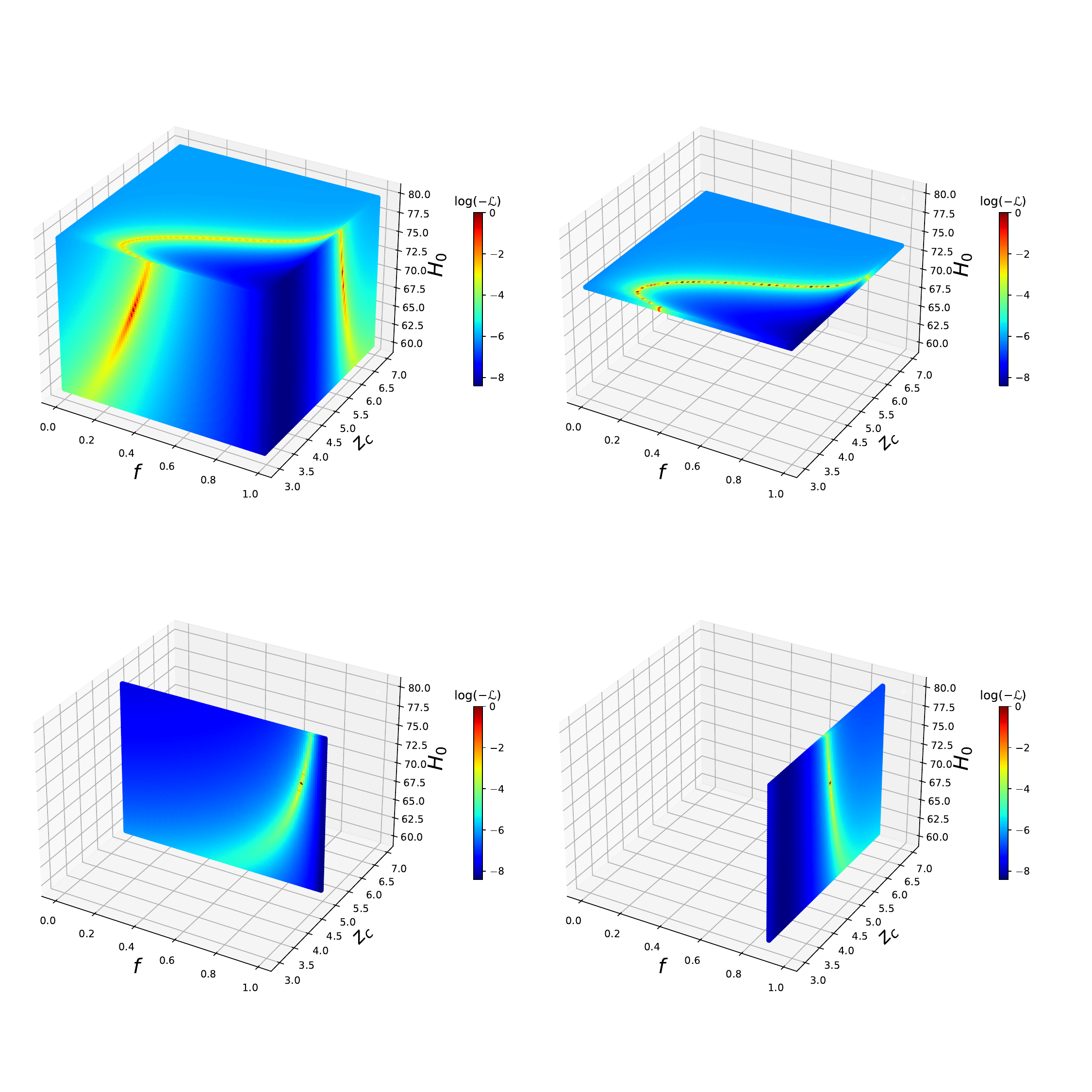}
    \caption{Likelihood over the parameter space for the EDE model, including slices for each parameter at the point of maximum Likelihood. The narrow path of high Likelihood shows the EDE model is effective at alleviating the Hubble tension for a very narrow range of parameters.}
    \label{fig:EDE_likelihood}
\end{figure}

Figure \ref{fig:EDE_likelihood} shows a plot of the Likelihood over the parameter space investigated ($H_0, f$ and $z_c$) in the MCMC analysis. The figure depicts the parameter space and slices through the space at the point of maximum Likelihood. It shows that, while the EDE model is effective at alleviating the Hubble tension, it is only effective for a very narrow range of parameters within that volume.

Using the python Dynesty package, the Bayesian evidence for this EDE model was $log(\cal{Z})$ = $-8.276$. This value, along with figure \ref{fig:EDE_likelihood}, indicates that EDE provides a good $\chi$-squared fit to the model but requires fine-tuned parameters and therefore is a highly specific model. $\Lambda$CDM, by comparison, is a good general model, effective over a large range of parameters.

\vskip 0.5cm

\begin{figure}[H]
    \centering
    \captionsetup{justification=justified, margin=1.5cm, font=small}
    \includegraphics[width=15.5cm]{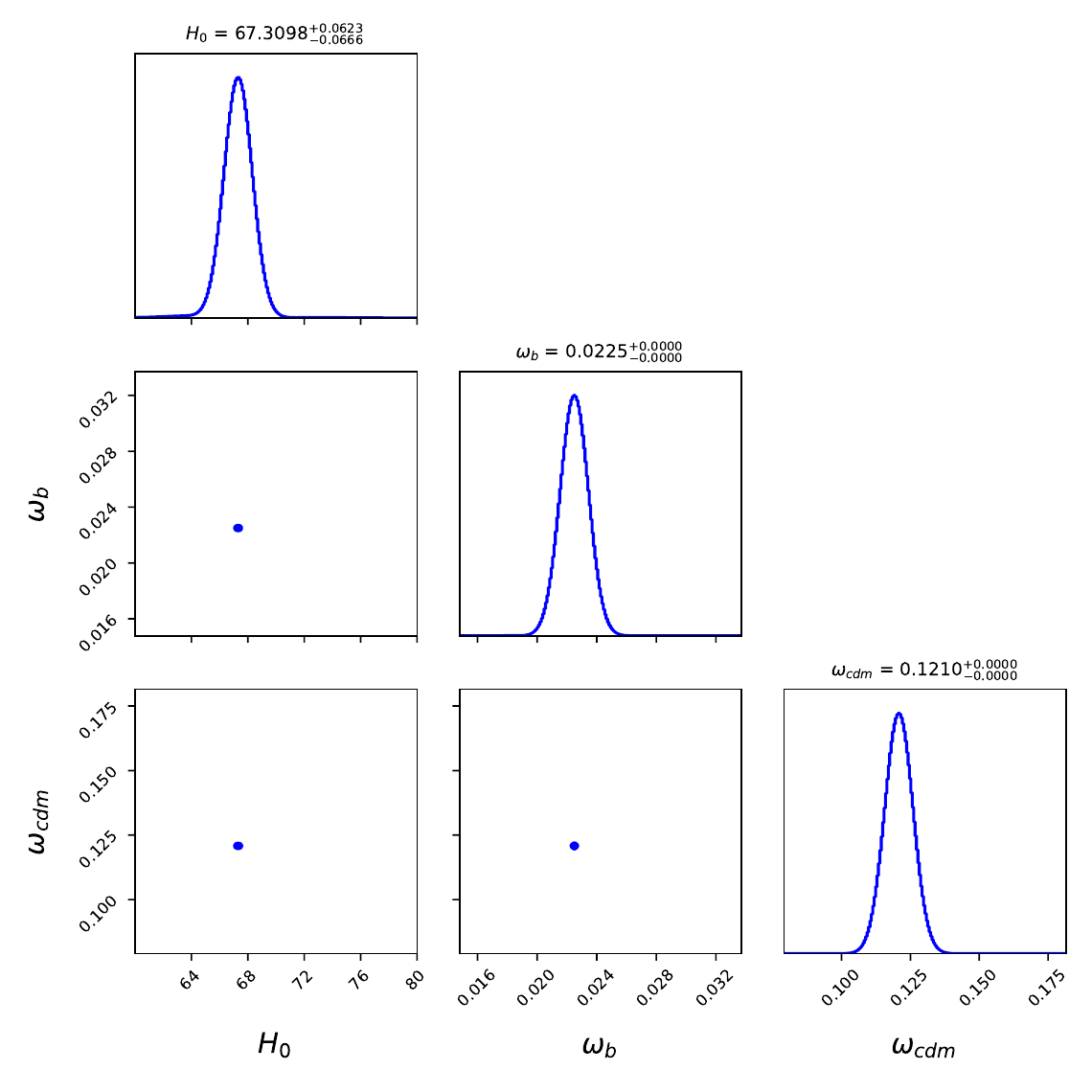}
    \caption{$H_0$ MCMC plot with no EDE and using Planck and Pantheon constraints. The values of $H_0, \omega_b$ and $\omega_{cdm}$ are in agreement with data obtained from the Planck Collaboration \cite{2020A&A...641A...6P} and are well constrained.}
    \label{fig:Lambda_mcmc}
\end{figure}

We compared the EDE model to a $\Lambda$CDM model without EDE and using constraints based on Planck 2018 results and $D_A$ based on Pantheon results \cite{2022ApJ...938..110B}. Figure \ref{fig:Lambda_mcmc} shows the MCMC analysis resulting in $H_0 = 67.31$ km s$^{-1}$ Mpc$^{-1}$, $\omega_b = 0.0225$ and $\omega_{cdm} = 0.1210.$ The results are in agreement with data obtained from the Planck Collaboration \cite{2020A&A...641A...6P}, confirming the validity of our model.

\begin{figure}[H]
    \centering
    \captionsetup{justification=justified, margin=1.5cm, font=small}
    \includegraphics[width=15.5cm]{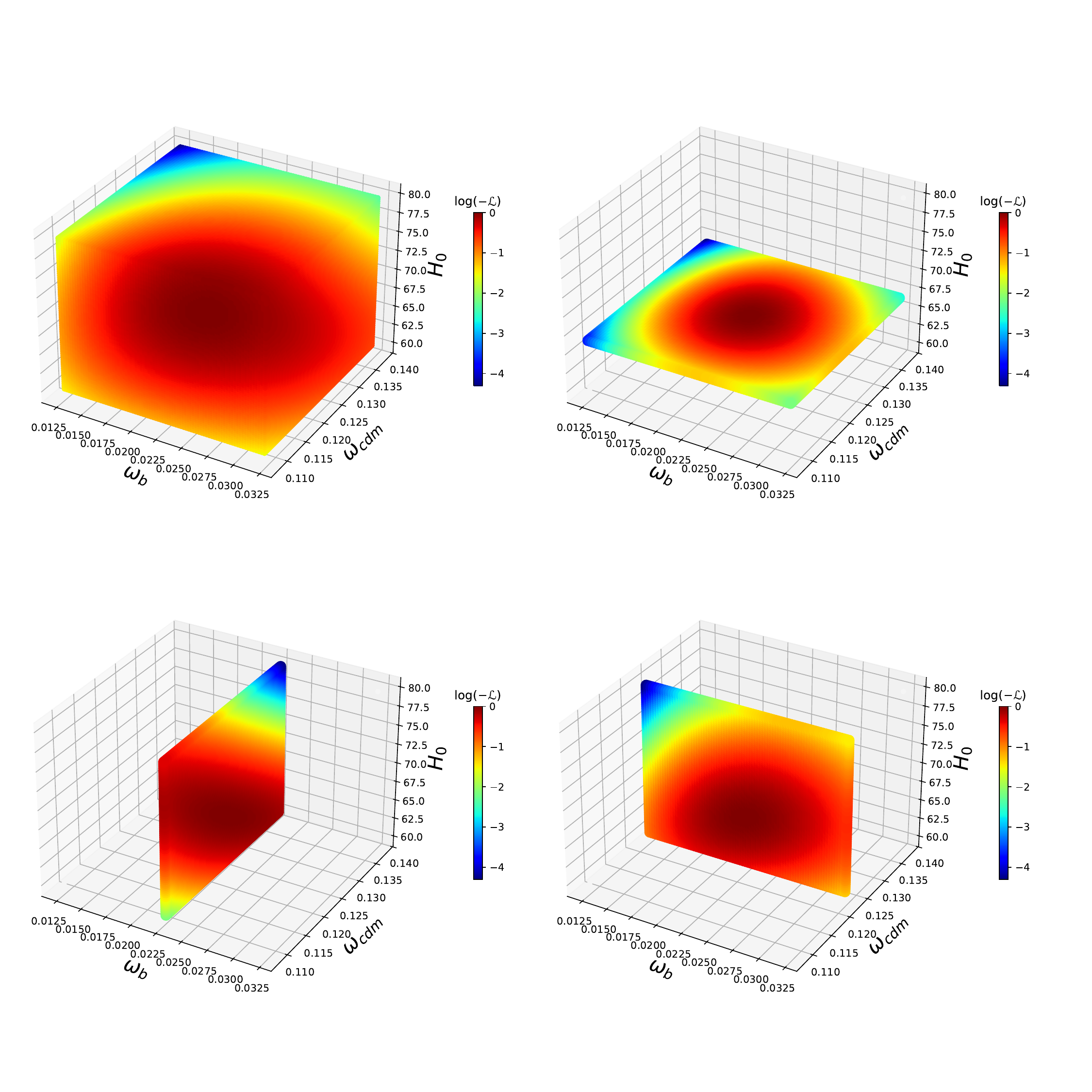}
    \caption{Likelihood cube for the $\Lambda$CDM model, based on Planck and Pantheon constraints. The high likelihood across a large volume indicates the model is effective for a wide range of parameters across the volume.}
    \label{fig:LCDM_Planck_Pantheon_likelihood}
\end{figure}

 Figure \ref{fig:LCDM_Planck_Pantheon_likelihood} shows a plot of the Likelihood over the parameter space investigated ($H_0, \omega_b$ and $\omega_{cdm}$) in the $\Lambda$CDM (Planck and Pantheon constraints) MCMC analysis. It shows that the model is effective for a wide range of parameters across the volume. The figure depicts the parameter space and slices through the space at the point of maximum Likelihood.

For this model, the Bayesian evidence was $log(\cal{Z})$ = $ -2269.147$ which indicates a less favoured model than our EDE model. However, the likelihood in figure \ref{fig:LCDM_Planck_Pantheon_likelihood} shows this model is favourable across a wider parameter space than the EDE model and does not require finely tuned conditions.

\begin{figure}[H]
    \centering
    \captionsetup{justification=justified, margin=2cm, font=small}
    \includegraphics[width=15.5cm]{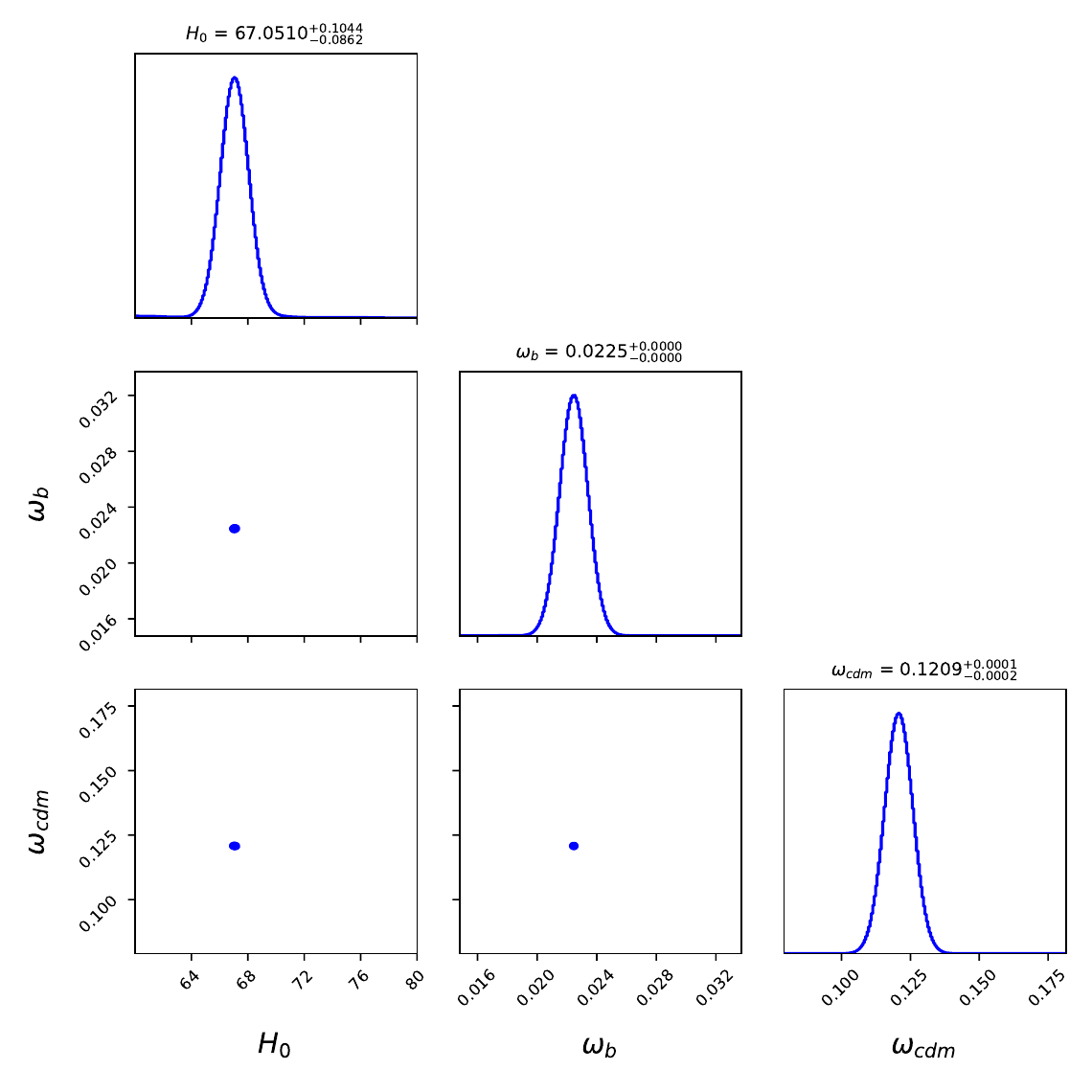}
    \caption{$H_0$ MCMC plot with no EDE and using only Planck constraints. Similar to figure \ref{fig:Lambda_mcmc}, the values of $H_0, \omega_b$ and $\omega_{cdm}$ are in agreement with data obtained from the Planck Collaboration \cite{2020A&A...641A...6P} and are tightly constrained.}
    \label{fig:Lambda_mcmc_Planck_only}
\end{figure}

\vskip 0.5cm
Lastly, we compared the EDE model to a $\Lambda$CDM model without EDE using constraints based on Planck 2018 results only. Figure \ref{fig:Lambda_mcmc_Planck_only} shows the results of the analysis, with $H_0 = 67.05$ km s$^{-1}$ Mpc$^{-1}$, $\omega_b = 0.0225$ and $\omega_{cdm} = 0.1209$, similar to the model based on Planck and Pantheon constraints.

\begin{figure}[H]
    \centering
    \captionsetup{justification=justified, margin=1.5cm, font=small}
    \includegraphics[width=15.5cm]{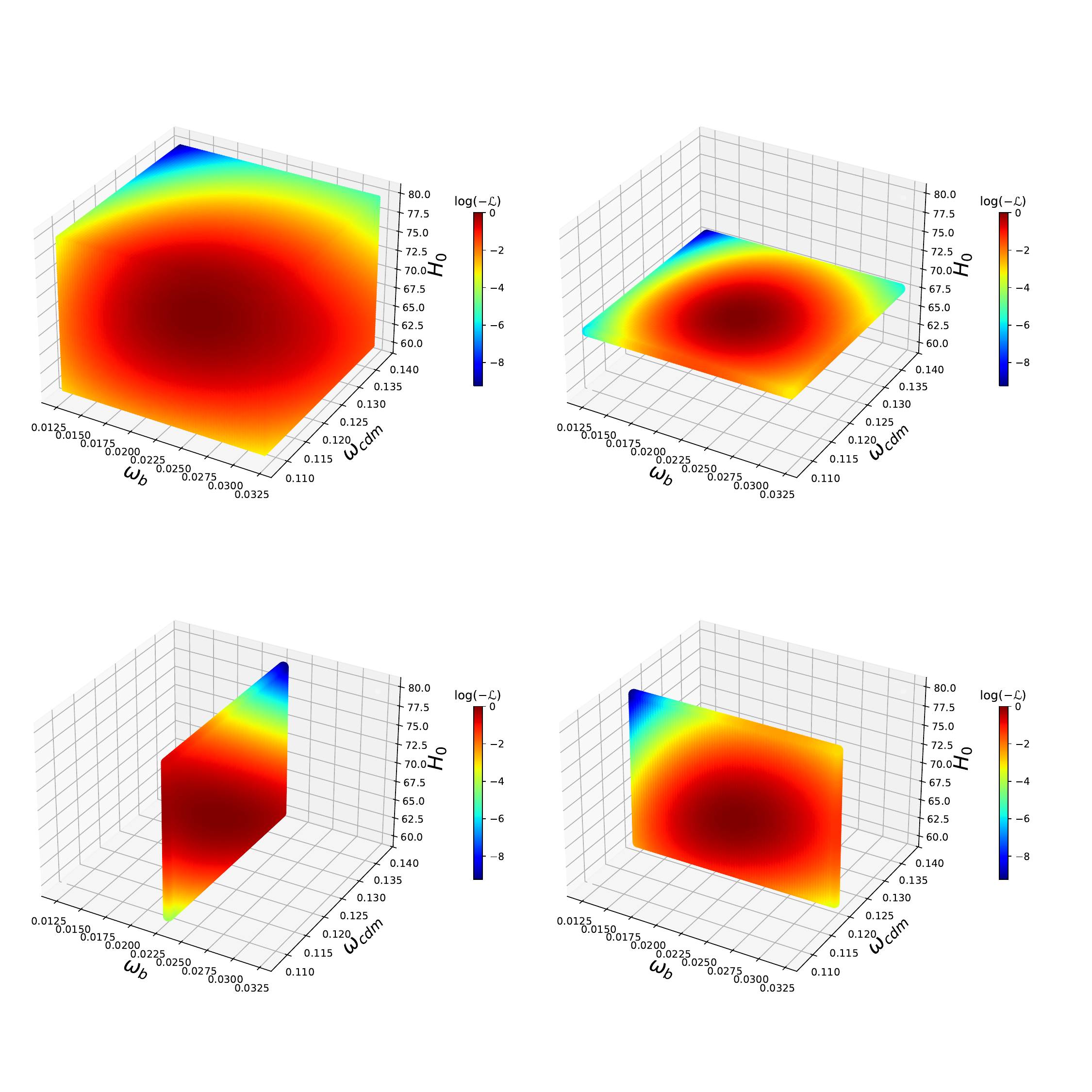}
    \caption{Likelihood cube for the $\Lambda$CDM model, based on Planck constraints only. This model shows a  slightly higher likelihood across a marginally wider volume than figure \ref{fig:LCDM_Planck_Pantheon_likelihood}, and is the model with the highest likelihood over a wide parameter space.}
    \label{fig:LCDM_Planck_only_likelihood}
\end{figure}

Figure \ref{fig:LCDM_Planck_only_likelihood} shows the Likelihood over the parameter space investigated ($H_0, \omega_b$ and $\omega_{cdm}$) in the $\Lambda$CDM (Planck and constraints only) MCMC analysis. This model shows a similar likelihood to figure \ref{fig:LCDM_Planck_Pantheon_likelihood}, but with a slightly higher likelihood across a marginally wider volume, and is the model with the highest likelihood over a wide parameter space.
The Bayesian evidence for this was $log(\cal{Z})$ = $ -23.975$, several orders of magnitude higher than the Planck/Pantheon constraints model.

\newpage
\section{Conclusion}\label{Conclusion}
While observations of the early universe are only possible as far back as the CMB, various theoretical models have been proposed to vary the state of the pre-CMB universe to resolve the Hubble tension.
In this paper we have reviewed a number of papers that each modified the conditions in the early universe, reducing the sound horizon and raising the value of $H_0$.
We have taken a phenomenological approach to the conditions of the early universe, exploring the effect of Early Dark Energy in the pre-CMB universe on the size of the sound horizon and its effect on the Hubble Constant.

Using an MCMC analysis, we found that a proportion of EDE can increase the Hubble constant to values consistent with measurements from the local universe. While effective, this method is a blunt approach and tends to move the onset of EDE to higher and higher redshifts, with EDE an increasing fraction of the universe and having a likelihood over a very narrow parameter space, particularly when compared to $\Lambda$CDM models. The median EDE fraction was found to be as high as $f = 0.798\ ^{+0.15}_{-0.28}$, at a median critical redshift $z_c = 60467 \ ^{+95043}_{-40204}$, resulting in $H_0 = 73.56$ km s$^{-1}$ Mpc$^{-1}$, which is in general agreement with observations from the local universe, such as SH0ES. This leaves questions as to the precise redshift and fraction that should occur for best effect. 

The Bayesian evidence for our EDE model, with $log(\cal{Z})$ = $-8.276$, and the plot of likelihood over the parameter space, indicates our EDE model is favoured for a narrow volume requiring highly fine-tuned parameters and therefore is not a good general model. The $\Lambda$CDM model based on Planck data has a slightly lower evidence, $log(\cal{Z})$ = $ -23.975$, but the plot of likelihood indicates this is a valid model over a wide range of parameters. 

We have shown that EDE can resolve the tension but only for a model with finely-tuned parameters. If the Hubble tension is resolved by a model just using EDE, our study shows that the parameters involving the distribution of EDE over time and the fraction of the universe are highly specific. If this small parameter space can be ruled out in some way, then we agree that EDE by itself will not resolve the Hubble tension and any model that does resolve the Hubble tension would most likely include physics of the early and late universe.

\vskip 1.5cm
\section*{Acknowledgments}
We would like to acknowledge the use of the high performance computing facilities provided by Digital Research Services, IT Services at the University of Tasmania.
We also acknowledge use of the following packages used in this paper: CAMB \cite{2011ascl.soft02026L}, emcee \cite{2013PASP..125..306F}, colossus.cosmology \cite{2018ApJS..239...35D}, corner package \cite{corner}, numpy \cite{harris2020array}, scipy \cite{2020SciPy-NMeth}, Matplotlib \cite{Hunter:2007} and Dynesty \cite{2020MNRAS.493.3132S, 2024zndo..12537467K, 2004AIPC..735..395S, 10.1214/06-BA127, 2009MNRAS.398.1601F}.

\bibliographystyle{plain}
\include{biby}

\end{document}

%% file: biby.tex
\bibliography{references}

%% file: main.bbl
\begin{thebibliography}{10}

\bibitem{2024Univ...10..305A}
{\"O}zg{\"u}r {Akarsu}, Eoin {{\'O} Colg{\'a}in}, Anjan~A. {Sen}, and M.~M. {Sheikh-Jabbari}.
\newblock {{\ensuremath{\Lambda}}CDM Tensions: Localising Missing Physics through Consistency Checks}.
\newblock {\em Universe}, 10(8):305, July 2024.

\bibitem{2023arXiv230200067B}
Ido {Ben-Dayan} and Utkarsh {Kumar}.
\newblock {Emergent Unparticles Dark Energy can restore cosmological concordance}.
\newblock {\em arXiv e-prints}, page arXiv:2302.00067, January 2023.

\bibitem{2023arXiv230705917B}
H.~B. {Benaoum}, Luz~{\'A}ngela {Garc{\'\i}a}, and Leonardo {Casta{\~n}eda}.
\newblock {Early dark energy induced by non-linear electrodynamics}.
\newblock {\em arXiv e-prints}, page arXiv:2307.05917, July 2023.

\bibitem{2022ApJ...938..110B}
Dillon {Brout}, Dan {Scolnic}, Brodie {Popovic}, Adam~G. {Riess}, et~al.
\newblock {The Pantheon+ Analysis: Cosmological Constraints}.
\newblock {\em The Astrophysical Journal}, 938(2):110, October 2022.

\bibitem{2025arXiv250314454C}
Erminia {Calabrese}, J.~Colin {Hill}, Hidde~T. {Jense}, et~al.
\newblock {The Atacama Cosmology Telescope: DR6 Constraints on Extended Cosmological Models}.
\newblock {\em arXiv e-prints}, page arXiv:2503.14454, March 2025.

\bibitem{2023arXiv230209091C}
Mariana {Carrillo Gonz{\'a}lez}, Qiuyue {Liang}, Jeremy {Sakstein}, and Mark {Trodden}.
\newblock {Neutrino-Assisted Early Dark Energy is a Natural Resolution of the Hubble Tension}.
\newblock {\em arXiv e-prints}, page arXiv:2302.09091, February 2023.

\bibitem{2023JHEP...06..052C}
Michele {Cicoli}, Matteo {Licheri}, Ratul {Mahanta}, Evan {McDonough}, Francisco~G. {Pedro}, and Marco {Scalisi}.
\newblock {Early Dark Energy in Type IIB String Theory}.
\newblock {\em Journal of High Energy Physics}, 2023(6):52, June 2023.

\bibitem{2023PhRvD.108b3518C}
Juan~S. {Cruz}, Steen {Hannestad}, Emil~Brinch {Holm}, Florian {Niedermann}, Martin~S. {Sloth}, and Thomas {Tram}.
\newblock {Profiling cold new early dark energy}.
\newblock {\em Physical Review D}, 108(2):023518, July 2023.

\bibitem{2023arXiv230204644D}
Diogo H.~F. {de Souza} and Rogerio {Rosenfeld}.
\newblock {Can neutrino-assisted early dark energy models ameliorate the $H_0$ tension in a natural way?}
\newblock {\em arXiv e-prints}, page arXiv:2302.04644, February 2023.

\bibitem{2018ApJS..239...35D}
Benedikt {Diemer}.
\newblock {COLOSSUS: A Python Toolkit for Cosmology, Large-scale Structure, and Dark Matter Halos}.
\newblock {\em The Astrophysical Journal Supplement Series}, 239(2):35, December 2018.

\bibitem{2023arXiv230714802E}
Luis~A. {Escamilla}, William {Giar{\`e}}, Eleonora {Di Valentino}, Rafael~C. {Nunes}, and Sunny {Vagnozzi}.
\newblock {The state of the dark energy equation of state circa 2023}.
\newblock {\em arXiv e-prints}, page arXiv:2307.14802, July 2023.

\bibitem{2023arXiv230315369E}
Johannes~R. {Eskilt}, Laura {Herold}, Eiichiro {Komatsu}, Kai {Murai}, Toshiya {Namikawa}, and Fumihiro {Naokawa}.
\newblock {Constraint on Early Dark Energy from Isotropic Cosmic Birefringence}.
\newblock {\em arXiv e-prints}, page arXiv:2303.15369, March 2023.

\bibitem{2009MNRAS.398.1601F}
F.~{Feroz}, M.~P. {Hobson}, and M.~{Bridges}.
\newblock {MULTINEST: an efficient and robust Bayesian inference tool for cosmology and particle physics}.
\newblock {\em Monthly Notices of the Royal Astronomical Society}, 398(4):1601--1614, October 2009.

\bibitem{corner}
Daniel Foreman-Mackey.
\newblock corner.py: Scatterplot matrices in python.
\newblock {\em The Journal of Open Source Software}, 1(2):24, jun 2016.

\bibitem{2013PASP..125..306F}
Daniel {Foreman-Mackey}, David~W. {Hogg}, Dustin {Lang}, and Jonathan {Goodman}.
\newblock {emcee: The MCMC Hammer}.
\newblock {\em Publications of the Astronomical Society of the Pacific}, 125(925):306, March 2013.

\bibitem{2023arXiv230300746G}
Samuel {Goldstein}, J.~Colin {Hill}, Vid {Ir{\v{s}}i{\v{c}}}, and Blake~D. {Sherwin}.
\newblock {Canonical Hubble-Tension-Resolving Early Dark Energy Cosmologies are Inconsistent with the Lyman-$\alpha$ Forest}.
\newblock {\em arXiv e-prints}, page arXiv:2303.00746, March 2023.

\bibitem{harris2020array}
Charles~R. Harris, K.~Jarrod Millman, et~al.
\newblock Array programming with {NumPy}.
\newblock {\em Nature}, 585(7825):357--362, September 2020.

\bibitem{2023PhRvD.108d3513H}
Laura {Herold} and Elisa G.~M. {Ferreira}.
\newblock {Resolving the Hubble tension with early dark energy}.
\newblock {\em Physical Review D}, 108(4):043513, August 2023.

\bibitem{2013ApJS..208...19H}
G.~{Hinshaw}, D.~{Larson}, E.~{Komatsu}, D.~N. {Spergel}, C.~L. {Bennett}, J.~{Dunkley}, M.~R. {Nolta}, M.~{Halpern}, R.~S. {Hill}, N.~{Odegard}, L.~{Page}, K.~M. {Smith}, J.~L. {Weiland}, B.~{Gold}, N.~{Jarosik}, A.~{Kogut}, M.~{Limon}, S.~S. {Meyer}, G.~S. {Tucker}, E.~{Wollack}, and E.~L. {Wright}.
\newblock {Nine-year Wilkinson Microwave Anisotropy Probe (WMAP) Observations: Cosmological Parameter Results}.
\newblock {\em The Astrophysical Journal Supplement}, 208(2):19, October 2013.

\bibitem{2023Univ....9...94H}
Jian-Ping {Hu} and Fa-Yin {Wang}.
\newblock {Hubble Tension: The Evidence of New Physics}.
\newblock {\em Universe}, 9(2):94, February 2023.

\bibitem{Hunter:2007}
J.~D. Hunter.
\newblock Matplotlib: A 2d graphics environment.
\newblock {\em Computing in Science \& Engineering}, 9(3):90--95, 2007.

\bibitem{2022PhRvD.105j3514J}
Jun-Qian {Jiang} and Yun-Song {Piao}.
\newblock {Toward early dark energy and n$_{s}$=1 with P l a n c k , ACT, and SPT observations}.
\newblock {\em Physical Review D}, 105(10):103514, May 2022.

\bibitem{2023ARNPS..73..153K}
Marc {Kamionkowski} and Adam~G. {Riess}.
\newblock {The Hubble Tension and Early Dark Energy}.
\newblock {\em Annual Review of Nuclear and Particle Science}, 73:153--180, September 2023.

\bibitem{2023PDU....3901170K}
Mohsen {Khodadi} and Marco {Schreck}.
\newblock {Hubble tension as a guide for refining the early Universe: Cosmologies with explicit local Lorentz and diffeomorphism violation}.
\newblock {\em Physics of the Dark Universe}, 39:101170, February 2023.

\bibitem{2024zndo..12537467K}
Sergey {Koposov}, Josh {Speagle}, Kyle {Barbary}, Gregory {Ashton}, Ed~{Bennett}, Johannes {Buchner}, Carl {Scheffler}, Ben {Cook}, Colm {Talbot}, James {Guillochon}, Patricio {Cubillos}, Andr{\'e}s {Asensio Ramos}, Matthieu {Dartiailh}, {Ilya}, Erik {Tollerud}, Dustin {Lang}, Ben {Johnson}, {jtmendel}, Edward {Higson}, Thomas {Vandal}, Tansu {Daylan}, Ruth {Angus}, {patelR}, Phillip {Cargile}, Patrick {Sheehan}, Matt {Pitkin}, Matthew {Kirk}, Joel {Leja}, {joezuntz}, and Danny {Goldstein}.
\newblock {joshspeagle/dynesty: v2.1.4}, June 2024.

\bibitem{2022arXiv221116394K}
V.~E. {Kuzmichev} and V.~V. {Kuzmichev}.
\newblock {The Hubble tension from the standpoint of quantum cosmology}.
\newblock {\em arXiv e-prints}, page arXiv:2211.16394, November 2022.

\bibitem{2023PhRvL.130p1003L}
Nanoom {Lee}, Yacine {Ali-Ha{\"\i}moud}, Nils {Sch{\"o}neberg}, and Vivian {Poulin}.
\newblock {What It Takes to Solve the Hubble Tension through Modifications of Cosmological Recombination}.
\newblock {\em Physical Review Letters}, 130(16):161003, April 2023.

\bibitem{2023EPJC...83..495L}
Thais {Lemos}, Joel~C. {Ruchika}, Carvalho, and Jailson {Alcaniz}.
\newblock {Low-redshift estimates of the absolute scale of baryon acoustic oscillations}.
\newblock {\em European Physical Journal C}, 83(6):495, June 2023.

\bibitem{2011ascl.soft02026L}
Antony {Lewis} and Anthony {Challinor}.
\newblock {CAMB: Code for Anisotropies in the Microwave Background}.
\newblock Astrophysics Source Code Library, record ascl:1102.026, February 2011.

\bibitem{2023PhRvD.107j3523L}
Meng-Xiang {Lin}, Evan {McDonough}, J.~Colin {Hill}, and Wayne {Hu}.
\newblock {Dark matter trigger for early dark energy coincidence}.
\newblock {\em Physical Review D}, 107(10):103523, May 2023.

\bibitem{2023arXiv230703481N}
Florian {Niedermann} and Martin~S. {Sloth}.
\newblock {New Early Dark Energy as a solution to the $H_0$ and $S_8$ tensions}.
\newblock {\em arXiv e-prints}, page arXiv:2307.03481, July 2023.

\bibitem{2023PhLB..84337988O}
Sergei~D. {Odintsov}, V.~K. {Oikonomou}, and German~S. {Sharov}.
\newblock {Early dark energy witPlanck Collaborationh power-law F(R) gravity}.
\newblock {\em Physics Letters B}, 843:137988, August 2023.

\bibitem{2021NuPhB.96615377O}
Sergei~D. {Odintsov}, Diego {S{\'a}ez-Chill{\'o}n G{\'o}mez}, and German~S. {Sharov}.
\newblock {Analyzing the H$_{0}$ tension in F(R) gravity models}.
\newblock {\em Nuclear Physics B}, 966:115377, May 2021.

\bibitem{2020A&A...641A...6P}
{Planck Collaboration} et~al.
\newblock {Planck 2018 results. VI. Cosmological parameters}.
\newblock {\em Astronomy \& Astrophysics}, 641:A6, September 2020.

\bibitem{2023arXiv230209032P}
Vivian {Poulin}, Tristan~L. {Smith}, and Tanvi {Karwal}.
\newblock {The Ups and Downs of Early Dark Energy solutions to the Hubble tension: a review of models, hints and constraints circa 2023}.
\newblock {\em arXiv e-prints}, page arXiv:2302.09032, February 2023.

\bibitem{2023PDU....4201348P}
Vivian {Poulin}, Tristan~L. {Smith}, and Tanvi {Karwal}.
\newblock {The Ups and Downs of Early Dark Energy solutions to the Hubble tension: A review of models, hints and constraints circa 2023}.
\newblock {\em Physics of the Dark Universe}, 42:101348, December 2023.

\bibitem{2022ApJ...934L...7R}
Adam~G. {Riess}, Wenlong {Yuan}, Lucas~M. {Macri}, Dan {Scolnic}, Dillon {Brout}, Stefano {Casertano}, David~O. {Jones}, Yukei {Murakami}, Gagandeep~S. {Anand}, Louise {Breuval}, Thomas~G. {Brink}, Alexei~V. {Filippenko}, Samantha {Hoffmann}, Saurabh~W. {Jha}, W.~{D'arcy Kenworthy}, John {Mackenty}, Benjamin~E. {Stahl}, and WeiKang {Zheng}.
\newblock {A Comprehensive Measurement of the Local Value of the Hubble Constant with 1 km s$^{-1}$ Mpc$^{-1}$ Uncertainty from the Hubble Space Telescope and the SH0ES Team}.
\newblock {\em The Astrophysical Journall}, 934(1):L7, July 2022.

\bibitem{2004AIPC..735..395S}
John {Skilling}.
\newblock {Nested Sampling}.
\newblock In Rainer {Fischer}, Roland {Preuss}, and Udo~Von {Toussaint}, editors, {\em Bayesian Inference and Maximum Entropy Methods in Science and Engineering: 24th International Workshop on Bayesian Inference and Maximum Entropy Methods in Science and Engineering}, volume 735 of {\em American Institute of Physics Conference Series}, pages 395--405. AIP, November 2004.

\bibitem{10.1214/06-BA127}
John Skilling.
\newblock {Nested sampling for general Bayesian computation}.
\newblock {\em Bayesian Analysis}, 1(4):833 -- 859, 2006.

\bibitem{2020MNRAS.493.3132S}
Joshua~S. {Speagle}.
\newblock {DYNESTY: a dynamic nested sampling package for estimating Bayesian posteriors and evidences}.
\newblock {\em Monthly Notices of the Royal Astronomical Society}, 493(3):3132--3158, April 2020.

\bibitem{2023arXiv230600454T}
Tomo {Takahashi} and Yo~{Toda}.
\newblock {Impact of big bang nucleosynthesis on the H0 tension}.
\newblock {\em arXiv e-prints}, page arXiv:2306.00454, June 2023.

\bibitem{2023PhRvD.107j3507T}
S.~X. {Tian} and Zong-Hong {Zhu}.
\newblock {Gravitation with modified fluid Lagrangian: Variational principle and an early dark energy model}.
\newblock {\em Physical Review D}, 107(10):103507, May 2023.

\bibitem{2023Univ....9..393V}
Sunny {Vagnozzi}.
\newblock {Seven Hints That Early-Time New Physics Alone Is Not Sufficient to Solve the Hubble Tension}.
\newblock {\em Universe}, 9(9):393, August 2023.

\bibitem{2020SciPy-NMeth}
Pauli Virtanen, Ralf Gommers, Travis~E. Oliphant, et~al.
\newblock {{SciPy} 1.0: Fundamental Algorithms for Scientific Computing in Python}.
\newblock {\em Nature Methods}, 17:261--272, 2020.

\bibitem{2022arXiv220909685W}
Hao {Wang} and Yun-Song {Piao}.
\newblock {A fraction of dark matter faded with early dark energy ?}
\newblock {\em arXiv e-prints}, page arXiv:2209.09685, September 2022.

\bibitem{2022PhLB..83237244W}
Hao {Wang} and Yun-Song {Piao}.
\newblock {Testing dark energy after pre-recombination early dark energy}.
\newblock {\em Physics Letters B}, 832:137244, September 2022.

\bibitem{2023arXiv230518873Y}
Gen {Ye}, Jun-Qian {Jiang}, and Yun-Song {Piao}.
\newblock {Shape of CMB lensing in the early dark energy cosmology}.
\newblock {\em arXiv e-prints}, page arXiv:2305.18873, May 2023.

\end{thebibliography}
